\documentclass[sigconf, authorversion, nonacm]{acmart}

\usepackage{gensymb}
\usepackage{amsmath}

\usepackage{subfig}
\usepackage{tabularx}
\usepackage{booktabs}
\usepackage{float}

\usepackage{balance} 
\usepackage[skip=0pt]{caption}

\AtBeginDocument{%
  }

\copyrightyear{2024} 
\acmYear{2024} 
\setcopyright{acmlicensed}\acmConference[ISWC '24]{Proceedings of the 2024 ACM International Symposium on Wearable Computers}{October 5--9, 2024}{Melbourne, VIC, Australia} \acmBooktitle{Proceedings of the 2024 ACM International Symposium on Wearable Computers (ISWC '24), October 5--9, 2024, Melbourne, VIC, Australia} 
\acmDOI{10.1145/3675095.3676619} 
\acmISBN{979-8-4007-1059-9/24/10}

\begin{document}

\newcommand{\name}{MunchSonic}{}

\newcommand{\add}[1]{\textcolor{black}{{#1}}}
\newcommand{\delete}[1]{{\sout{ #1}}}


\title{\name{}: Tracking Fine-grained Dietary Actions through Active Acoustic Sensing on Eyeglasses}


\author{Saif Mahmud}
\affiliation{%
  \institution{Cornell University}
  \city{Ithaca, NY}
  \country{USA}}
\email{sm2446@cornell.edu}
\orcid{0000-0002-5283-0765}

\author{Devansh Agarwal}
\affiliation{%
  \institution{Cornell University}
  \city{Ithaca, NY}
  \country{USA}}
\email{da398@cornell.edu}
\orcid{0009-0005-1338-9275}

\author{Ashwin Ajit}
\affiliation{%
  \institution{Cornell University}
  \city{Ithaca, NY}
  \country{USA}}
\email{aa794@cornell.edu}
\orcid{0009-0009-0700-688X}

\author{Qikang Liang}
\affiliation{%
  \institution{Cornell University}
  \city{Ithaca, NY}
  \country{USA}}
\email{ql75@cornell.edu}
\orcid{0009-0001-9301-625X}

\author{Thalia Viranda}
\affiliation{%
  \institution{Cornell Tech}
  \city{New York, NY}
  \country{USA}}
\email{tv74@cornell.edu}
\orcid{0009-0008-6933-3926}

\author{Francois Guimbretiere}
\affiliation{%
  \institution{Cornell University}
  \city{Ithaca, NY}
  \country{USA}}
\email{francois@cs.cornell.edu}
\orcid{0000-0002-5510-6799}

\author{Cheng Zhang}
\affiliation{%
  \institution{Cornell University}
  \city{Ithaca, NY}
  \country{USA}}
\email{chengzhang@cornell.edu}
\orcid{0000-0002-5079-5927}

\renewcommand{\shortauthors}{Saif Mahmud et al.}

\begin{abstract}
We introduce \name{}, an AI-powered active acoustic sensing system integrated into eyeglasses to track fine-grained dietary actions. \name{} emits inaudible ultrasonic waves from the eyeglass frame, with the reflected signals capturing detailed positions and movements of body parts, including the mouth, jaw, arms, and hands involved in eating. These signals are processed by a deep learning pipeline to classify six actions: hand-to-mouth movements for food intake, chewing, drinking, talking, face-hand touching, and other activities (null). In an unconstrained study with 12 participants, \name{} achieved a 93.5\% macro F1-score in a user-independent evaluation with a 2-second resolution in tracking these actions, also demonstrating its effectiveness in tracking eating episodes and food intake frequency within those episodes.
\end{abstract}



\maketitle

\section{Introduction}
\label{sec:intro}

Tracking dietary activity is crucial for measuring nutrition levels and managing chronic diseases. However, the lack of reliable automated dietary monitoring systems forces most people to rely on manual input-based applications, which are prone to self-reporting errors. To address this, researchers have developed wearable dietary activity recognition systems. Despite advancements in sensing modalities and machine learning algorithms to process the data, tracking fine-grained dietary actions such as food intake, chewing, and drinking, while distinguishing them from similar body movements like talking, or face touching, as well as from non-eating activities, remains a significant challenge for the wearable computing research community.

Dietary activities such as eating and drinking require coordinated hand, jaw, and mouth movements. Detecting fine-grained dietary actions necessitates the simultaneous tracking of both hand and mouth movements. Current state-of-the-art dietary monitoring systems, whether on eyeglasses~\cite{mydj, fitbyte, mirtchouk2017}, earables~\cite{auracle}, or smartwatches~\cite{eatingtrak}, require sensing modalities at multiple body locations to achieve this, which can be costly, power-inefficient, and cumbersome for many users. 

Recently, the active acoustic sensing system ActSonic~\cite{mahmud2024actsonic} showed promising efficacy in tracking everyday activities, including identifying eating and drinking episodes, using a single point of instrumentation on eyeglasses. However, this system has not yet been able to predict fine-grained dietary actions, such as chewing or hand-to-mouth intake gestures, which are related to important biomarkers for longitudinal health monitoring~\cite{slyper_oral_2021}, such as chewing patterns, drinking habits, or snacking behaviors. In this paper, we aim to address the following research question:

\begin{itemize}
    \item \textit{Can an active acoustic sensing system instrumented on smart glasses be used to further distinguish fine-grained dietary actions in unconstrained environments beyond only identifying eating moments?}
\end{itemize}


To address this challenge, we developed \name{}, a fine-grained dietary action recognition system in the form factor of commodity eyeglasses using active acoustic sensing. Similar to ActSonic~\cite{mahmud2024actsonic}, our system employs two pairs of microphones and speakers attached to the hinges of the eyeglasses. \name{} transmits inaudible ultrasonic waves that reflect off various body parts, generating a two-dimensional echo profile, which serves as a range spectrogram of the scanned body parts. This profile is then analyzed by a lightweight deep learning framework to infer fine-grained dietary actions. 

\add{\name{} is based on the observation that fine-grained dietary actions involve movements of the mouth, jaw, and arms, which can be inferred from reflected ultrasound, as shown in prior work~\cite{PoseSonic,eario,eyeEcho, SonicASL}. However, the capacity of the active acoustic sensing-based general-purpose activity recognition system ActSonic~\cite{mahmud2024actsonic} is limited to tracking moments of eating and drinking episodes. Since eating is a complex activity involving several fine-grained dietary actions such as hand-to-mouth movement for intake, chewing, swallowing, etc., we designed the \name{} system specifically for tracking body parts involved in eating, with modifications to ActSonic~\cite{mahmud2024actsonic} hardware and inference pipeline. We evaluated our system in a user study with 12 participants in unconstrained environments of their choice. The results showed that \name{} can accurately recognize fine-grained dietary actions such as hand movements for intake, chewing, and drinking, as well as non-dietary actions like talking, face touching, and other activities (null), achieving an F1-score of 93.5\%.}

The contributions of the \name{} system are as follows:

\begin{itemize}
    \item  Advanced the knowledge of wearable-based active acoustic sensing by demonstrating the feasibility of using active acoustic sensing on glasses to recognize fine-grained dietary actions beyond only identifying eating moments.
  
    \item Evaluation of the proposed system through a 12-participant user study in unconstrained environments, with ground truth data annotated for each second.
\end{itemize}

\section{Related Work}
\label{sec:rel-work}

\begin{table*}[ht!]
\fontsize{7pt}{7pt}\selectfont
\begin{tabular}{@{}|c|c|l|l|l|l|l|@{}}
\toprule
\textbf{Year} & \textbf{Study}   & \textbf{Wearable}                                                           & \textbf{\begin{tabular}[c]{@{}l@{}}Sensing\\ Modalitie(s)\end{tabular}}                             & \textbf{\begin{tabular}[c]{@{}l@{}}Detected\\ Event(s)\end{tabular}}             & \textbf{\begin{tabular}[c]{@{}l@{}}Evaluation\\ Metric\end{tabular}}                                                                       & \textbf{\begin{tabular}[c]{@{}l@{}}Power\\ Consumption\end{tabular}}       \\ \midrule
2017          & EarBit~\cite{bedri2017earbit}           & \begin{tabular}[c]{@{}l@{}}Outer Ear\\ Interface\end{tabular}               & \begin{tabular}[c]{@{}l@{}}IMU, IR Proxomity\\ Sensor, Microphone\end{tabular}                    & Chewing                                                                          & F1-score = 80.1\%                                                                                                                          & -                                                                          \\ \midrule
2017          & GlasSense~\cite{chung2017glasses}        & Eyeglasses                                                                  & Load Cell                                                                                         & \begin{tabular}[c]{@{}l@{}}Head Movement,\\ Talking, Chewing\end{tabular} & F1-score = 94.0\%$*$                                                                                                                         & -                                                                          \\ \midrule
2017          & Mirtchouk et al.~\cite{mirtchouk2017} & Google Glass                                                                & \begin{tabular}[c]{@{}l@{}}IMU, Microphone,\\ Motion Sensor\end{tabular}                        & Meals                                                                            & \begin{tabular}[c]{@{}l@{}}Precision = 31\%\\ Recall = 87\%\end{tabular}                                                                   & -                                                                          \\ \midrule
2018          & Auracle~\cite{auracle}          & Earpiece                                                                    & \begin{tabular}[c]{@{}l@{}}Contact\\ Microphone\end{tabular}                                      & \begin{tabular}[c]{@{}l@{}}Eating \\ Episodes\end{tabular}                       & F1-score = 77.5\%                                                                                                                          & \begin{tabular}[c]{@{}l@{}}14.47 mW\end{tabular} \\ \midrule
2018          & Zhang et al.~\cite{zhang2017monitoring}     & Eyeglasses                                                                  & EMG Electrode                                                                                     & \begin{tabular}[c]{@{}l@{}}Eating \\ Event\end{tabular}                          & F1-score = 77\%                                                                                                                            & 81.96 mW                                                                   \\ \midrule
2020          & FitByte~\cite{fitbyte}          & Eyeglasses                                                                  & \begin{tabular}[c]{@{}l@{}}1 Camera, \\ 1 Proximity\\ Sensor, 6 IMUs\end{tabular}                 & \begin{tabular}[c]{@{}l@{}}Eating and\\ Drinking\\ Episodes\end{tabular}         & \begin{tabular}[c]{@{}l@{}}Precision = 82.8\% (Eat- \\ -ing), 56.7\% (Drinking)\\ Recall = 93.8\% (Eating), \\ 65.5\% (Drinking)\end{tabular} & 105.08 mW                                                                  \\ \midrule
2022          & MyDJ~\cite{mydj}             & Eyeglasses                                                                  & \begin{tabular}[c]{@{}l@{}}Accelerometer, \\ Piezoelectric\end{tabular}                           & \begin{tabular}[c]{@{}l@{}}Eating\\ Episodes\end{tabular}                        & F1-score = 92\%                                                                                                                            & 26.06 mW                                                                   \\ \midrule
\textbf{2024}          & \textbf{\name{}}       & \textbf{Eyeglasses}                                                                  & \begin{tabular}[c]{@{}l@{}}\textbf{Active Acoustic} \\ \textbf{Sensing}\end{tabular}                                & \begin{tabular}[c]{@{}l@{}}\textbf{Intake, Chewing, Drink-}\\ \textbf{-ing, Talking, Face Touch}\end{tabular}                        & \textbf{F1-score = 93.50\%}                                                                                                                        & \textbf{96.5 mW\textsuperscript{\textdagger}}                                                                    \\ \bottomrule
\end{tabular}
\caption{\add{Comparison of eating detection studies based on wearable form factors, sensing modalities, event granularity, evaluation metrics (F1-score or precision/recall), and power draw (mW). Asterisks ($*$) in the \textit{Evaluation Metric} column indicate lab study evaluations; otherwise, evaluations were in free-living conditions. The dagger symbol (\textdagger) in the \textit{Power Consumption} column indicates machine learning inference was performed on a cloud or mobile device, not the wearable.}}
\label{table:prior-works}
\end{table*}

Various wearable devices with different sensing modalities have been proposed to track eating events. This section discusses non-eyeglass wearables, followed by eyeglasses. Table~\ref{table:prior-works} summarizes prior studies on tracking eating activities.

\textbf{Form Factors Other than Eyeglasses.} 
Smartwatch-based systems for eating~\cite{thomaz2015practical, ye2016assisting, dong2013detecting, kyritsis2019detecting, sen2018annapurna, eatingtrak, thomaz2015_02} and drinking~\cite{hamatani2018fluidmeter} use IMUs to track hand-to-mouth gestures but face challenges distinguishing similar motions, leading to high false positives~\cite{nic-lane-imu, chun2018detecting, ecai-har, kdd-har}. Ear-worn devices, or earables, track eating by monitoring jaw movement through in-ear proximity sensing~\cite{bedri2015, bedri2015wearable}, passive acoustic sensing~\cite{gao2016ihear, auracle}, or both~\cite{bedri2017earbit}. Neckbands~\cite{farooq2014novel, bodybeat, yatani2012bodyscope} and necklaces~\cite{alshurafa2015recognition, kalantarian2015monitoring, edison2018, shin2019accurate, necksense} use passive acoustic and proximity sensing to track chewing and swallowing. \add{Despite advancements and around $80\%$ accuracy in free-living conditions, neckband and necklace form factors have not gained the same popularity as other wearables~\cite{amores2019exploration}.}

\textbf{Eyeglasses Form Factor.}
Dietary monitoring using eyeglasses is promising due to their proximity to the mouth and jaw and higher social acceptance~\cite{AttentivU} (64\% adoption in the US~\cite{farooq2018accelerometer}). Researchers have used EMG electrodes~\cite{zhang2017monitoring}, piezoelectric sensors~\cite{farooq2016novel, farooq2016segmentation}, inertial sensors~\cite{farooq2018accelerometer}, and load cells~\cite{chung2017glasses} on eyeglasses to track eating behaviors. Sensor fusion approaches~\cite{fitbyte, rahman2016, mirtchouk2017, fitnibble, mydj}, integrating gyroscopes, accelerometers, proximity sensors, and microphones, have shown improved performance. Passive acoustic sensing using contact microphones~\cite{auracle} captures chewing sounds but requires close skin contact. Systems using acoustic sensing~\cite{mahmud2024actsonic, mollyn2022samosa} focus on eating episodes rather than fine-grained actions.

In summary, \name{} is the first to use active acoustic sensing on eyeglasses to infer fine-grained dietary actions by capturing jaw and hand movements. This allows for detailed dietary action extraction, such as intake and chewing, unlike other systems that primarily track moments of eating episodes.
\section{System Implementation}
\label{sec:implementation}

The goal of \name{} is to capture the information related to movements on multiple body parts involved in dietary activities using an eyeglasses form factor. Previous research has shown active acoustic sensing's efficacy on wearables in tracking facial muscle movements~\cite{eario, li2024eyeecho, echospeech, zhang2023hpspeech, EchoNose}, upper body limbs~\cite{PoseSonic}, hand poses~\cite{lee2024echowrist, yu2024ring}, gaze~\cite{GazeTrak}, respiration~\cite{c-fmcw}, sign language gestures~\cite{SonicASL, TransASL}, physiological signals~\cite{fan2023apg} and everyday activities~\cite{mahmud2024actsonic}. Inspired by these prior works, especially~\cite{mahmud2024actsonic}, \name{} integrates active acoustic sensing into eyeglasses to track these fine-grained dietary actions. This section discusses the active acoustic sensing mechanism, the hardware implementation, and the deep learning framework for processing the captured data.

\subsection{Sensing Mechanism}
\label{sec:sensing-mechanism} 
The design of \name{}'s sensing system is inspired by the active acoustic sensing approach from~\cite{mahmud2024actsonic}.  In essence, \name{} uses a similar active acoustic sensing processing method, based on cross-correlation-based Frequency Modulated Continuous Wave (C-FMCW)~\cite{c-fmcw}. These chirps are transmitted from eyeglasses' hinges, with frequency ranges of 18-21.5 KHz and 21.5-24.5 KHz for the left and right transmitters, respectively. The receiver samples at 50 KHz, with each chirp containing 600 samples and a sweep period of 0.012 seconds. The system can detect changes as small as 3.43 mm, given the slow speed of mouth, jaw, and hand movements compared to the speed of sound (343 m/s). The sensing range is up to 2.058 meters, though a shorter range is used for fine-grained dietary actions. \add{\name{} uses a lower amplifier gain compared to~\cite{mahmud2024actsonic} to focus on mouth, jaw, and hand movements near the face for tracking fine-grained dietary actions. Consequently, the microphones receive weaker reflections from lower body parts that are not involved in dietary actions.}

\name{}'s signal processing pipeline involves computing the cross-correlation of transmitted and received ultrasonic waves. The received signal is first filtered through a bandpass filter (18-21 KHz and 21.5-24.5 KHz) to eliminate audible frequencies. The cross-correlation output, called the \textit{Echo Profile}, functions as a Range-FFT using acoustic C-FMCW waves. The \textit{Echo Profile} is processed as a two-dimensional tensor ($x$-axis: time, $y$-axis: distance from \name{} eyeglasses form factor). To capture body part movements and eliminate static object reflections, the first derivative of the \textit{Echo Profile}, termed the \textit{Differential Echo Profile}, is used as input for the deep learning architecture to track fine-grained dietary actions.

\subsection{Hardware and Form Factor}
\label{sec:hardware}

\begin{figure}[!htbp]
    \centering
    \includegraphics[width=0.9\columnwidth]{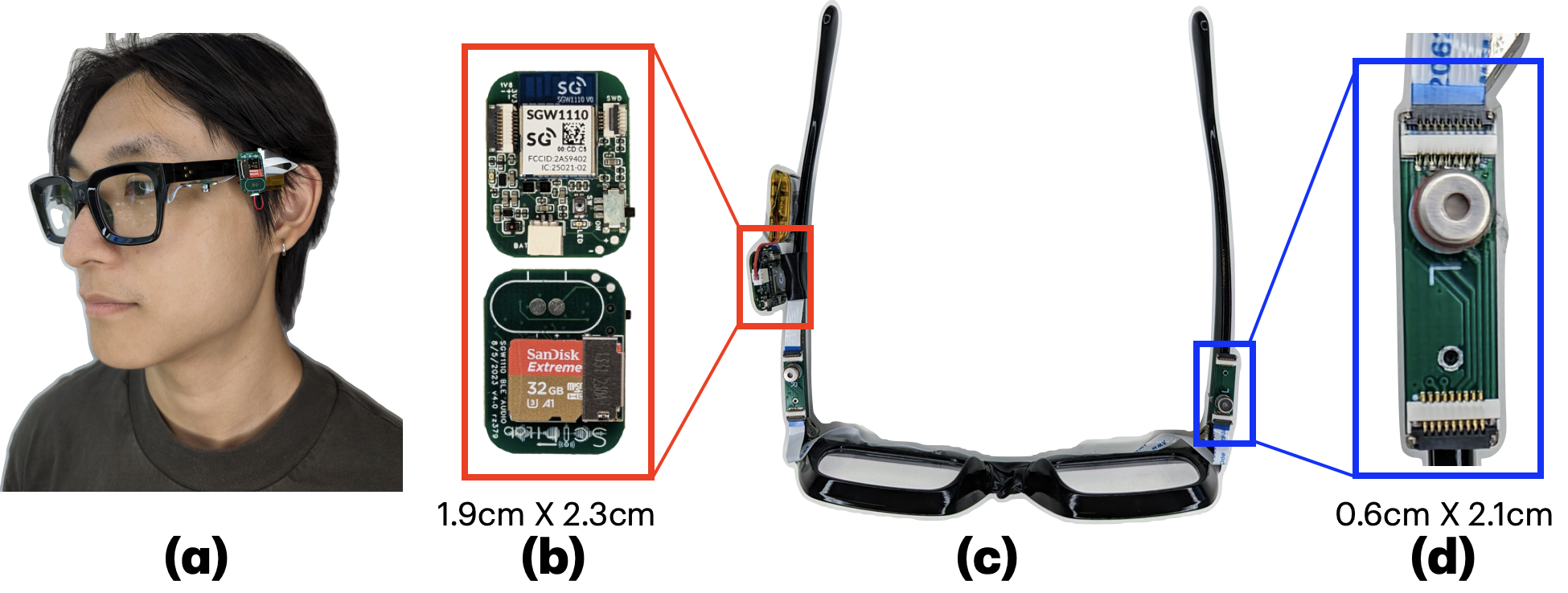}
    \caption{\name{} Hardware and Form Factor: (a) User wearing eyeglasses form factor, (b) Customized controller unit with nRF52840 microcontroller, (c) Top view of the eyeglasses form factor, (d) \name{} transceiver for active acoustic sensing housing one speaker (top) and one microphone (bottom).}
    \label{fig:hardware}
\end{figure}

We replicated the sensing system from ActSonic~\cite{mahmud2024actsonic} with minor changes to the sensors and controller unit, using the OWR-05049T-38D speaker and ICS-43434 microphone~\cite{mic-i2s}. Our customized controller unit, shown in Figure~\ref{fig:hardware}(b), features the nRF52840 microcontroller~\cite{nRF52840}, two MAX98357A audio amplifiers, a BLE SGW1110~\cite{ble-module} module, power management modules, and an SD card slot with a SanDisk Extreme Pro microSD card~\cite{sd-card}, optimized for power efficiency. The transceiver boards, each with a speaker and microphone, are attached to the eyeglass hinges, with placement refined for optimal tracking of mouth, jaw, and hand movements. The controller unit and LiPo battery are attached to one leg of the glasses frame, connected via Flexible Printed Circuit (FPC) cables and a JST connector. \add{The sensing system on the eyeglasses draws 96.5 mW of power for collecting data, with a voltage of 4.02V and a current of 24.0 mA. The prototype used in the \name{} study, featuring a 290 mAh LiPo battery, lasts approximately 11.25 hours in SD card storage mode.}

\subsection{Deep Learning Framework}
\label{sec:dl-model}

\subsubsection{Data Processing and Model Architecture}
We designed a lightweight deep learning framework to process the active acoustic data captured by the \name{} system. The differential echo profile described in Section~\ref{sec:sensing-mechanism} serves as the input to the model. We create overlapping sliding windows from the differential echo profile data. To augment the data, we apply Gaussian noise to 5\% of the windows and mask 5\% of the distance axis to simulate environmental noise and unrelated movements.

The preprocessed sliding windows are fed into a MobileNetV2~\cite{sandler2018mobilenetv2} convolutional neural network encoder, which generates a 256-dimensional embedding vector for each window. This vector is then fed into a classifier network with three feedforward layers (128, 64, and 6 neurons). Each layer, except the last, includes batch normalization~\cite{batch-norm}, dropout~\cite{srivastava2014dropout} with a probability of 0.25, and Leaky ReLU~\cite{xu2015empirical} activation. Finally, a softmax operation provides the class probability distribution.

\subsubsection{Model Training}
The input to \name{}'s deep learning framework consists of sliding windows of differential echo profiles, optimally 2 seconds long with 50\% overlap. This window length fits an eating intake gesture. The optimal sensing range is 150 pixels on the $y$-axis of the differential echo profile, corresponding to 51.45 cm from the \name{} device. With speakers and microphones on each hinge of the eyeglasses, transmitting acoustic waves in two frequency ranges, there are four channels in the differential echo profile. Thus, the shape of the input window is $(4 \times 150 \times 166)$, as each second contains 83 samples.

We use focal loss~\cite{lin2017focal} for training to handle class imbalance in the \name{} dataset. The Adam~\cite{kingma2014adam} optimizer with a cosine annealing learning rate scheduler starts with an initial learning rate of $10^{-2}$. The model, implemented using PyTorch and PyTorch Lightning, is trained for 30 epochs with a batch size of 128 on GeForce RTX 2080 Ti GPUs.
\section{User Study}
\label{sec:user-study}

We conducted a 45-minute user study to evaluate \name{} in an unconstrained setting. Building on ActSonic's~\cite{mahmud2024actsonic} success in tracking eating and drinking episodes with high F1 scores, the \name{} study assumed active acoustic sensing on glasses can distinguish eating moments from other activities. This paper focuses on extracting fine-grained dietary actions, such as hand-to-mouth movements, chewing, and drinking. The study aimed to assess the system's performance in tracking these actions and distinguishing them from non-eating activities in real-world conditions. We conducted a shorter-duration free-living study, leveraging ActSonic's proven efficacy in longer studies. Ground truth annotation was done per second, unlike other systems~\cite{mydj} that rely on self-reporting, enabling high-resolution recognition of dietary actions.

\textbf{Study Protocol} We conducted a user study approved by our organization's Institutional Review Board (IRB) with 13 participants, 6 identifying as male and 7 as female, with an average age of \(26.5\) years, ranging from 21 to 36 years. The participants wore the \name{} eyeglasses form factor and a chest-mounted GoPro HERO9 camera~\cite{gopro} facing upward to capture their eating activities around the face as the ground truth. The camera recorded 720p video at 30 fps with a diagonal field of view of \(148^\circ\).  Note that one participant (P09) accidentally turned off the chest-mounted camera, resulting in no ground truth data for that participant. Therefore, we discarded that participant from the study and evaluated the system on the remaining 12 participants.

\begin{figure}[!htbp]
    \centering
    \includegraphics[width=1.0\columnwidth]{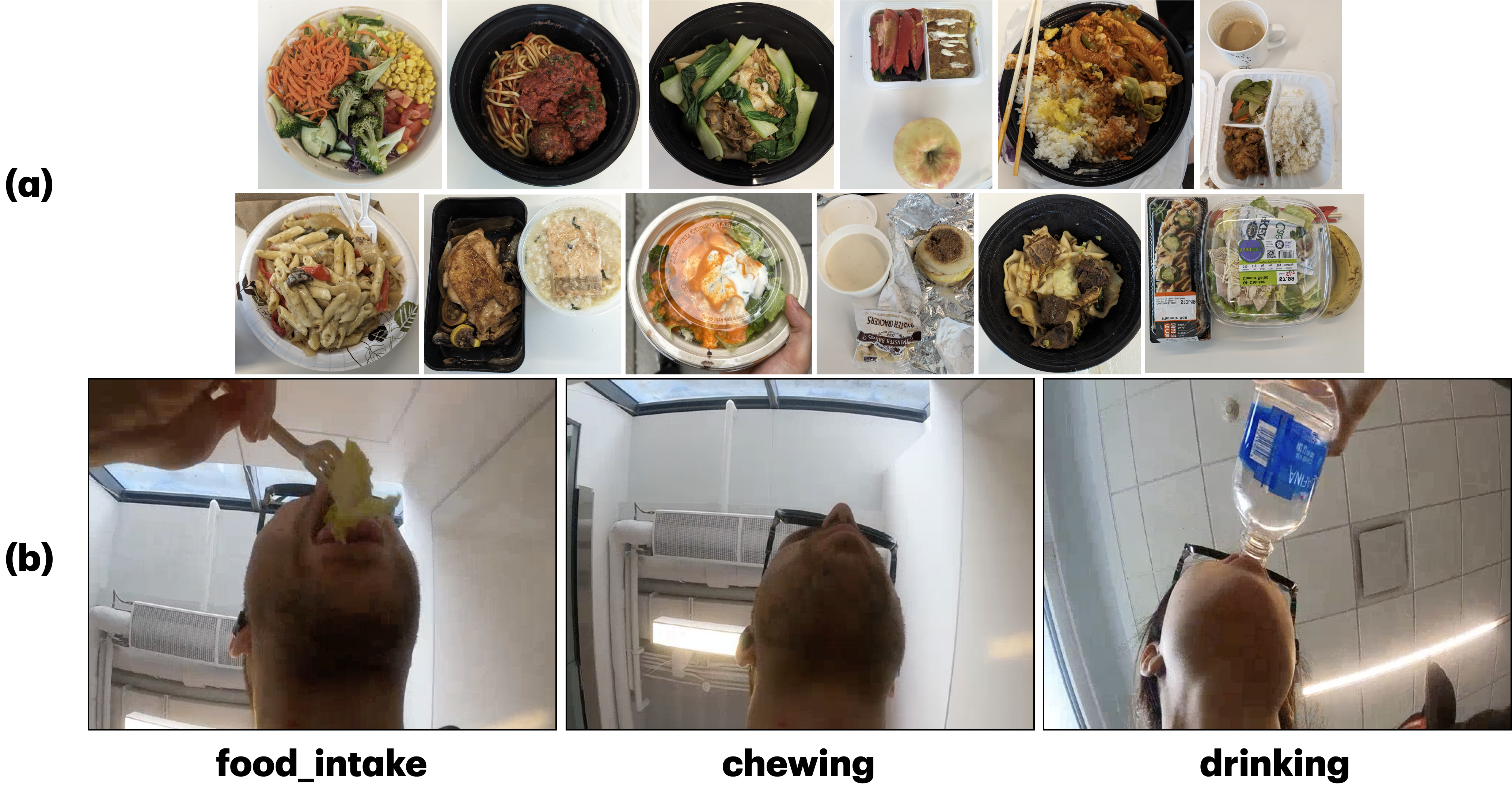}
    \caption{User study in unconstrained conditions: (a) Foods consumed by user study participants, \add{(b) Sample images of the chest-mounted camera view of \name{} data collection pipeline.}}
    \label{fig:user-study}
\end{figure}

Participants were asked to bring a meal (breakfast, lunch, or dinner) from any source (home-cooked or restaurant) and come to the lab. After briefing them about the study procedure, we equipped them with the \name{} eyeglasses and the chest-mounted camera. We verified the data collection system with a short 5-minute session. Once verified, the data collection system was started, and participants were free to go anywhere while wearing the eyeglasses and camera. The only requirement was to finish the meal they brought to ensure sufficient samples of dietary movements. Apart from that, they could continue with their daily routine. The total duration of this unconstrained study was 45 minutes. Upon completion, participants returned to the lab to return the devices and were reimbursed with a \$20 USD gift card.

\textbf{Dataset Statistics} \add{We collected 540 minutes of data from 12 participants. Two researchers labeled the reference video from the chest-mounted camera at each second using the ANU-CVML Video Annotation Tool (Vidat)~\cite{zhang2020vidat}. The inter-rater reliability was measured through Cohen's Kappa~\cite{lazar2017research}, $\kappa$, and its value is 0.82 ($\kappa > 0.80$ denotes near perfect agreement, and $0.6 < \kappa \le 0.8$ denotes satisfactory agreement). The camera view that the annotators labeled is shown in Figure~\ref{fig:user-study}(b). The data was categorized into six classes: hand-to-mouth movement for food or drink intake, chewing, drinking, talking, face touch, and a null class for other activities. The annotators were instructed to segment the actions in two phases. First, they determined whether the action was eating or non-eating. If the action was identified as eating, they further categorized it into fine-grained actions such as hand-to-mouth movement, chewing, and drinking. Swallowing was integrated into the chewing category. For non-eating actions, they segmented the activities into talking, face touching, and other activities not included in the tracking set. A synchronization script was used to extrapolate ground truth labels for the differential echo profile sliding windows. The class distribution was 53.5\% null, 5.9\% hand-to-mouth for food intake, 22.0\% chewing, 2.2\% drinking, 15.1\% talking, and 1.3\% touching face with hand for non-eating activities.}
\section{Performance Evaluation}
\label{sec:eval}

We computed precision, recall, and macro F1-score to assess \name{}'s ability to track fine-grained dietary actions. Evaluations were conducted at two levels: frame-level inference for each two-second sliding window and episode-level inference for intake count and chewing time estimation. Employing a leave-one-participant-out evaluation strategy enabled user-independent assessment, using data from one participant as the test set and the remaining 11 for training and validation. This evaluation demonstrates the system's performance without requiring training data from new users, which makes the system easier to deploy at scale.

\subsection{Evaluation of Frame-Level Inference}

\begin{figure}[!htbp]
    \centering
    \includegraphics[width=0.70\columnwidth]{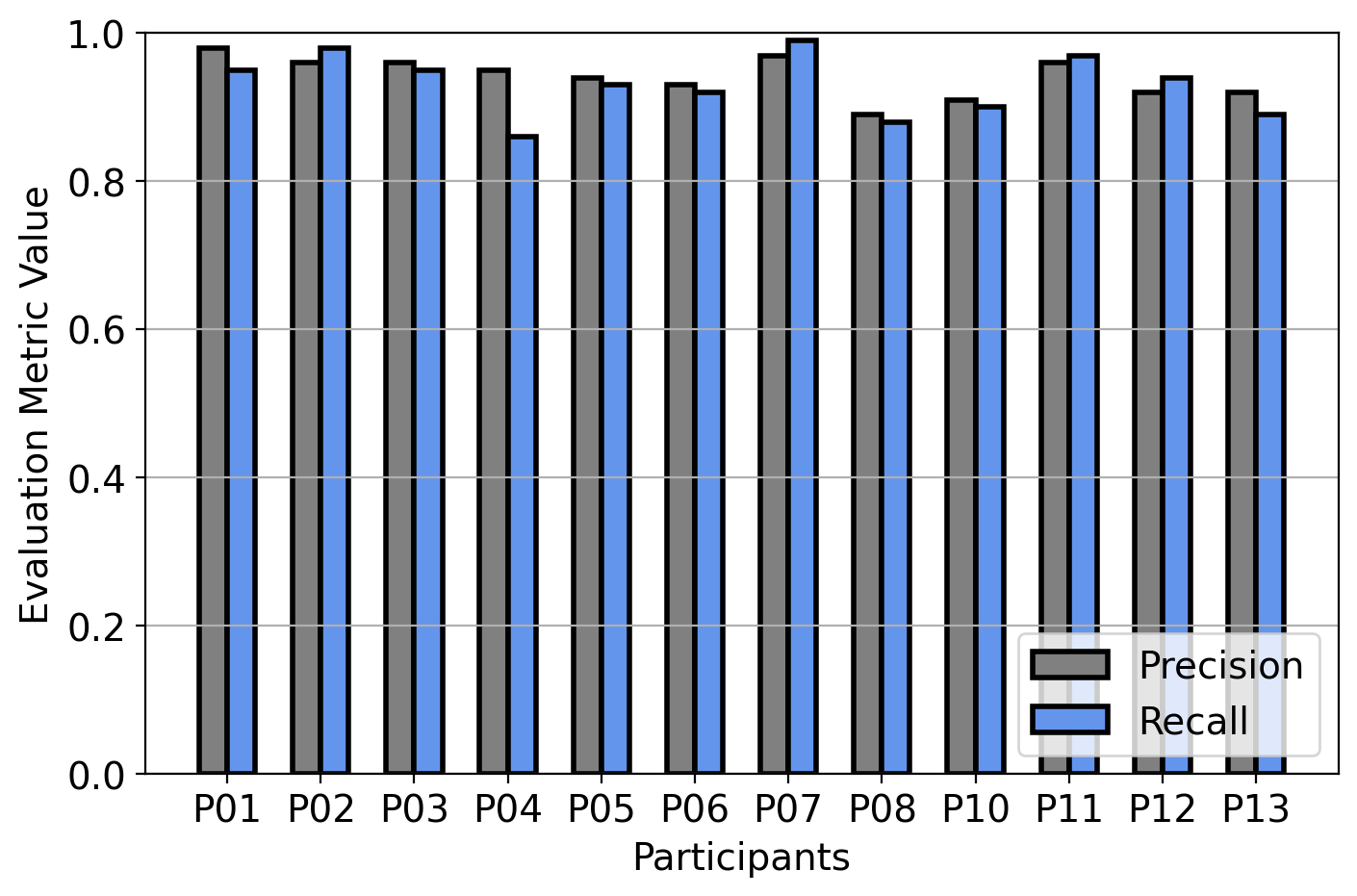}
    \caption{Precision and recall of leave-one-participant-out evaluation of \name{}, where data from each participant on the $x$-axis serves as the test set.}
    \label{fig:part-wise}
\end{figure}

\begin{figure}[!htbp]
    \centering
    \includegraphics[width=0.9\columnwidth]{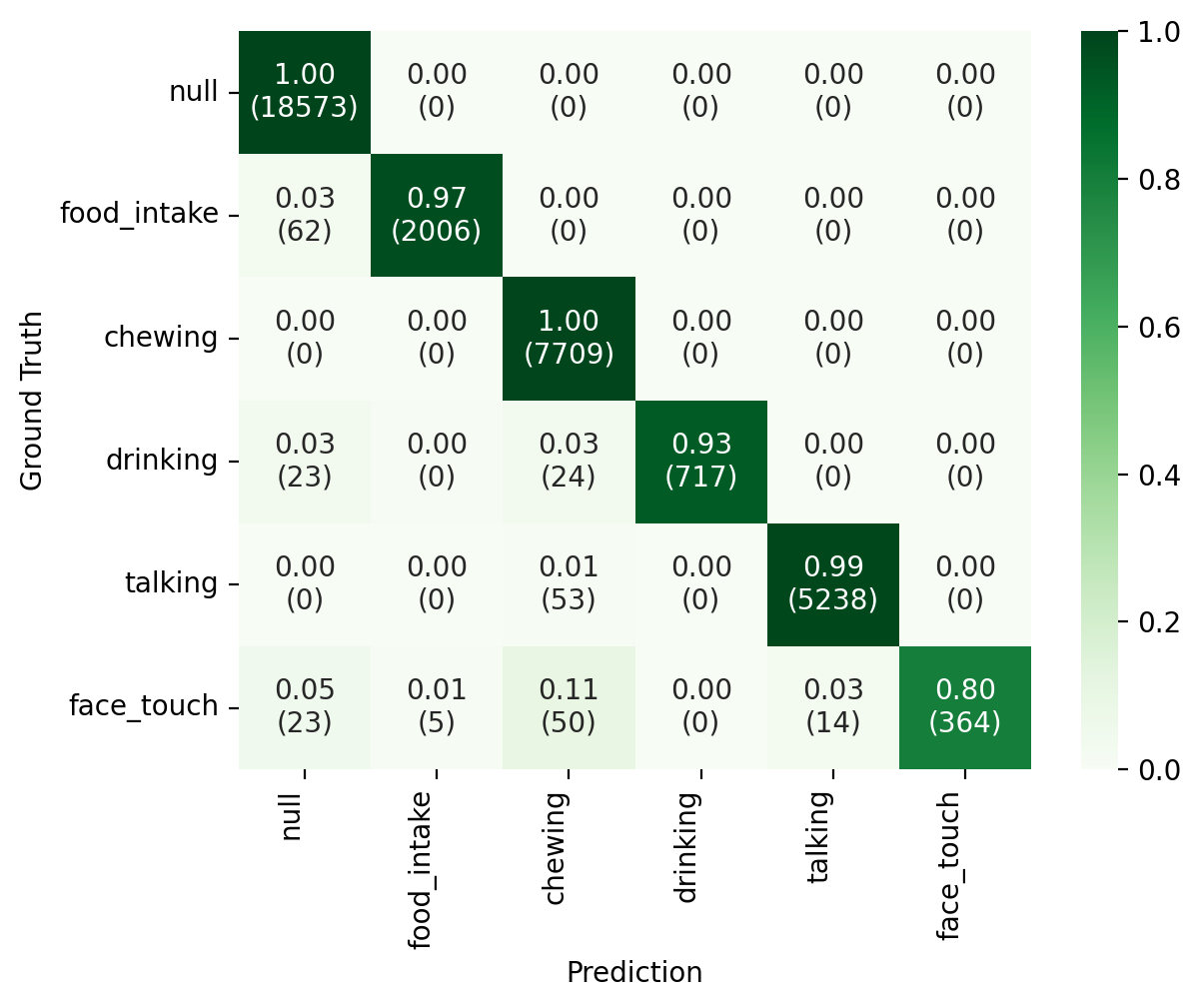}
    \caption{Normalized confusion matrix from the leave-one-participant-out evaluation across 12 participants in the user study of \name{}. \add{The values in parentheses represent the total number of instances for each cell.}}
    \label{fig:conf-mat}
\end{figure}

We summarize the results of the leave-one-participant-out evaluation of sliding window prediction in Figure~\ref{fig:part-wise}. According to the evaluation, the average macro F1-score across all 12 participants is 0.935 with a standard deviation of 0.031. It is evident that the \name{} system can track fine-grained dietary events with precision and recall of more than 90\% for most participants. P08 demonstrates the worst performance in terms of both precision and recall. Additionally, P04 and P13 yielded slightly lower recall compared to the average. Our analysis of the ground truth videos for those participants suggests that the slightly worse performance can be attributed to the participants touching their faces several times during the study for activities not related to eating. This phenomenon is also evident in the confusion matrix in Figure~\ref{fig:conf-mat}. The performance degradation of \name{} for face-touch actions can be related to the imbalanced nature of the dataset collected in a completely unconstrained manner, which contains very few samples of this particular activity. Nonetheless, \name{} exhibits robust performance across all activities, with a mean precision and recall of $0.941 \pm 0.027$ and $0.930 \pm 0.041$ respectively, across all 12 participants.

\subsection{Evaluation of Episode-Level Inference}
\begin{figure}[!htbp]
    \centering
    \includegraphics[width=\columnwidth]{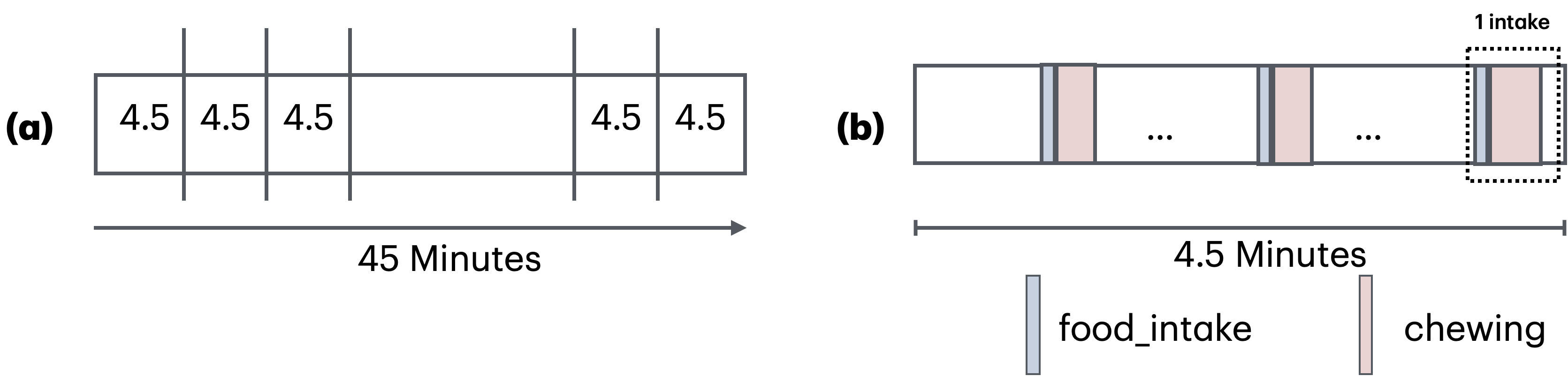}
    \caption{Episode-level evaluation of \name{}: (a) Segmentation of user study data into 4.5-minute-long episodes, (b) Detection of food intake within each episode.}
    \label{fig:episode-eval}
\end{figure}

\subsubsection{Eating Episode Detection}
To evaluate \name{} in detecting eating episodes and intake counts within episodes, we segmented the 45-minute data collected from each participant in the user study into 10 segments of equal length (4.5 minutes each). According to the definition provided in~\cite{thomaz2015practical, eatingtrak}, if a segment of length $t_w$ contains $\frac{t_w + 20}{5}$ intakes, which is 5 intakes for the aforementioned segments, then that segment is defined as an eating episode. To evaluate \name{}'s efficacy in detecting these eating episodes, we deployed a majority voting strategy where a segment is labeled as an eating episode if 50\% of the frames are labeled as either \textit{food\_intake} or \textit{chewing}. The number of undetected eating episodes for the \name{} dataset is 1 out of 43 ground truth episodes, resulting in a False Negative Rate of 0.023. Additionally, there were 77 non-eating episodes in the dataset, of which 2 were incorrectly detected as eating by \name{}, leading to a False Positive Rate of 0.026.

\subsubsection{Food Intake Counting}
Furthermore, to count the number of intakes in each segment, we increment the intake count for that segment by 1 if we find one \textit{food\_intake} frame followed by at least two chewing frames or windows within the next 3.0 seconds. We compute the Mean Absolute Error (MAE) between the ground truth number of intakes and the predicted count as a metric to evaluate this. The average ground truth intake count in one 4.5-minute eating segment (containing more than 5 ground truth intakes within the timeframe) and non-eating segment (containing fewer than 5 ground truth intakes) is 17.65 and 1.19, respectively. The MAE for counting intakes in the eating segments is 2.01, leading to an average error of 11.39\% in predicting the number of eating intakes. For the non-eating segments, the MAE for counting intakes is 0.047, leading to a mean error of 3.95\% in predicting the intake counts during non-eating episodes.

\subsubsection{Chewing Time Estimation}
To evaluate the coverage of predicted chewing time, we calculate the Mean Absolute Error (MAE) between the total ground truth chewing time in a 4.5-minute episode and the predicted chewing time, measured in seconds, across all 12 user study participants. The average chewing time across eating episodes is 163.6 seconds (2.73 minutes) and 9.17 seconds for non-eating episodes in the \name{} dataset. Interpolating from the frame-level predictions of \name{}, the MAE of the estimated chewing time for eating episodes is 12.87 seconds, leading to a coverage of 92.13\% of chewing time. For the non-eating episodes, the MAE of the estimated chewing time is 1.61 seconds, leading to a coverage of 82.5\%. This higher coverage across non-eating episodes indicates \name{}'s potential in detecting snacking events, which was outside the scope of the user study.

\section{Discussion}
\label{sec:discussion}

\begin{figure*}[!htbp]
    \centering
    \includegraphics[width=0.75\textwidth]{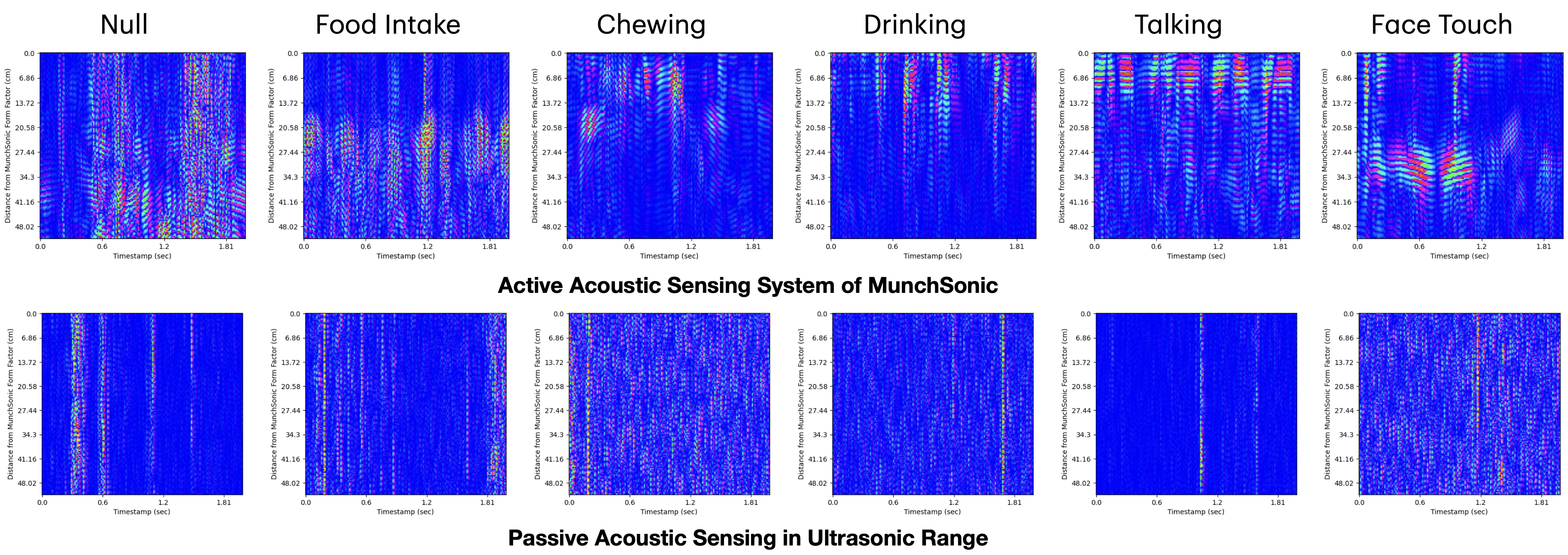}
    \caption{Differential echo profiles of actions with active and passive acoustic sensing. The $x$-axis represents time, and the $y$-axis represents the echo distance from the \name{} form factor, ranging from 0 cm to 50 cm.}
    \label{fig:echo-profiles}
\end{figure*}

\textbf{Analyzing the Passive Sensed Ultrasonic Range for Tracking Dietary Actions} Theoretically, our system focuses on nearly inaudible sound frequencies above 18 KHz, which should not interfere with most daily activities, including eating. However, some eating sounds can have frequency components above 18 KHz. To evaluate the impact of these passively sensed ultrasonic signals on distinguishing dietary actions, we conducted a preliminary study with one researcher as a participant. In this study, we turned off \name{}'s speakers to prevent them from transmitting C-FMCW chirps and performed the same set of activities as in the user study. The microphones continued to record the surrounding acoustic signals. We then created differential echo profile windows and trained the \name{} deep learning model. The mean cross-session macro F1-score for this user-dependent model was 0.322. In contrast, with the speakers transmitting ultrasonic chirps, the mean cross-session macro F1-score was 0.976. As illustrated in Figure~\ref{fig:echo-profiles}, active acoustic sensing plays a crucial role in distinguishing dietary activities by tracking the movements of body parts involved in dietary actions. While passive ultrasonic signals might have a minor impact on detecting fine-grained events, quantifying this impact numerically is challenging and beyond the scope of this paper. Our experiments indicate that chewing crunchy foods generated more ultrasonic components in the spectrogram. However, the signal strength of reflections from \name{}'s transmitted ultrasonic chirps is much higher, potentially overshadowing the passively sensed components' feature importance. Further investigation into this will be left for future work.

\begin{figure}[!htbp]
    \centering
    \includegraphics[width=0.75\columnwidth]{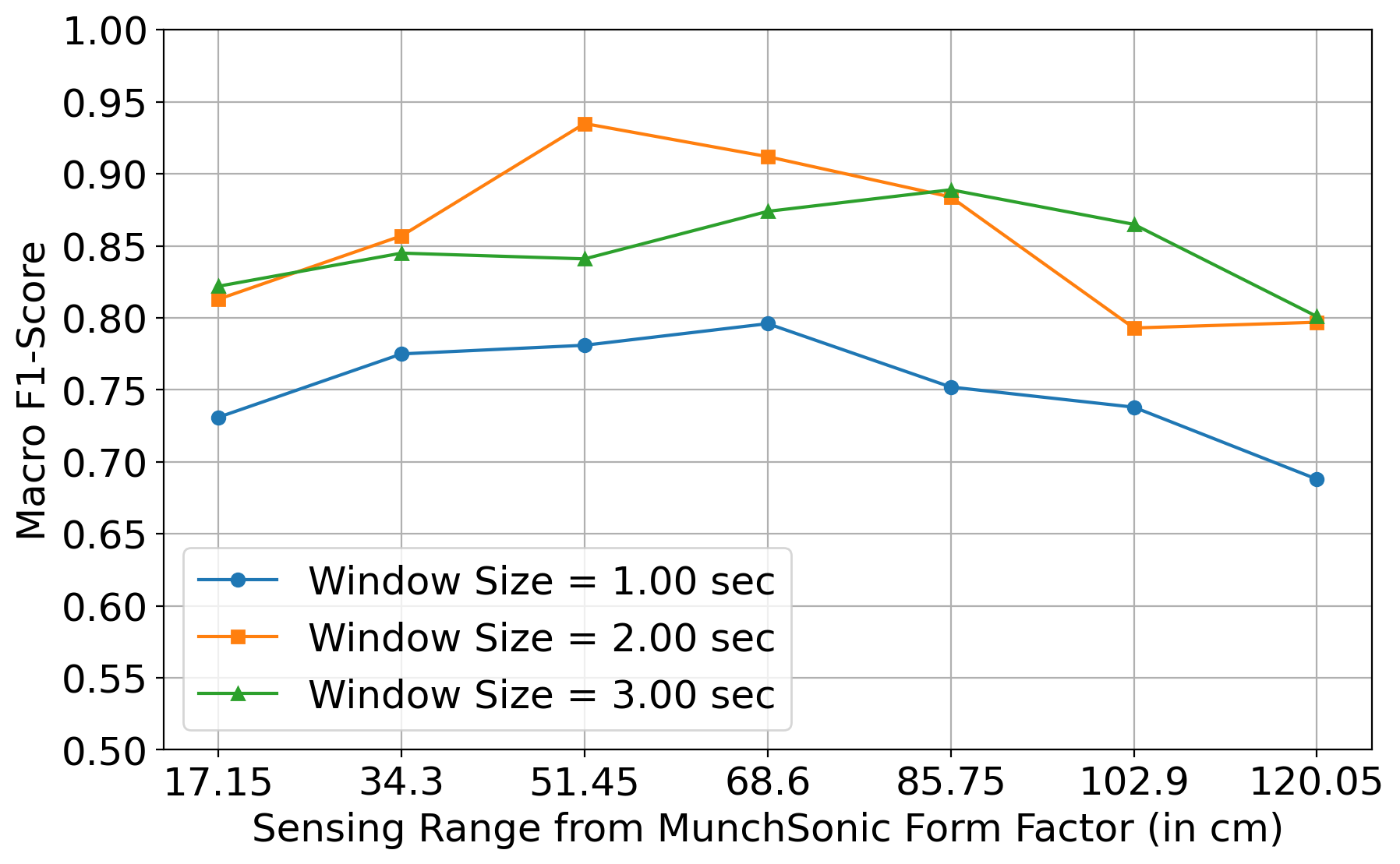}
    \caption{Ablation study to measure the impact of sensing range and sliding window size on the performance of \name{}. Here, all the sliding window sizes mentioned have a 50\% overlap.}
    \label{fig:ablation}
\end{figure}

\textbf{Impact of Sensing Range and Temporal Context} We evaluated the impact of sensing range and temporal context for the \name{} system. Figure~\ref{fig:ablation} shows that a sliding window size of 2.00 seconds with a 50\% overlap yields the best tracking performance. We also found that a sensing range of 50 to 80 cm is optimal for tracking fine-grained dietary actions. This is because most movements related to dietary actions occur in the jaw and mouth region, within 30 cm of the eyeglasses form factor. Additionally, \name{} tracks hand movements for food intake, so extending the sensing range to 50 cm yields the best performance. Poorer performance at longer ranges is likely due to various unrelated activities occurring in that region. 

\textbf{Comfort and Safety of Wearing \name{} Prototype} \add{After participating in the user study described in Section~\ref{sec:user-study}, participants completed an IRB-approved questionnaire on the comfort of wearing the \name{} prototype. They rated their comfort from 0 (most uncomfortable) to 5 (most comfortable), with a mean rating of 3.462 and a standard deviation of 1.050. Since \name{} operates using the ultrasonic range, participants were not expected to hear anything from the system, and indeed, none reported hearing any noise. Three participants noted that the prototype did not fit their head sizes and prescription lenses well, which could be addressed in future iterations by creating an attachable sensing module. Additionally, one participant mentioned that the chest-mounted camera used for ground truth acquisition occasionally touched their dining table, interrupting their eating.}

\add{We measured the transmitted signal intensity using a CDC-provided app~\cite{cdc-app}, finding it to be 68 dB(A), below the NIOSH limit of 85 dB~\cite{murphy2002revisiting}. While MHz range ultrasonic exposure can cause discomfort~\cite{moyano2022possible}, \name{} operates in the KHz range with no reported issues. Future studies will explore potential audibility among animals and children.}

\textbf{Potential Application Scenarios} We envision \name{} being useful for eating behavior assessment, chronic disease management, and overall well-being. Our system passively and objectively measures eating episodes and micro-level actions (e.g., pauses, drinking, chewing) in real-world settings. By accurately classifying these actions, we can derive meal micro-structure metrics~\cite{bellisle_edograms_2020} like eating duration, speed, and chewing rate, traditionally assessed through interviews or self-report questionnaires~\cite{sasaki_self-reported_2003}, which are prone to biases. Our automated approach provides precise, naturalistic data. Clinical studies link these metrics to health outcomes like higher Body Mass Index~\cite{sonoda_associations_2018, walsh_eating_1989}, eating disorders~\cite{fairburn_binge_1993, ede_17}, and metabolic diseases~\cite{yuan_association_2021}. Accurate assessment enhances monitoring and management of these conditions, facilitating real-time interventions. A day-long evaluation of \name{} will provide crucial insights into snacking behavior and bulimia or binge eating disorder detection.

\section{Conclusion}
\name{} accurately tracks fine-grained dietary actions with high precision and scalability in real-world settings, surpassing other systems and advancing dietary health assessment and intervention design.

\section*{Acknowledgements}
This project was supported by the National Science Foundation Grant No. 2239569 and partially by the Cornell University IGNITE Innovation Acceleration Program.

\bibliographystyle{ACM-Reference-Format}
\balance
\bibliography{main}

\end{document}